\documentstyle[12pt]{article}
\makeatletter
\@addtoreset{equation}{section}
\makeatother

\begin{document}
%
 \mbox{} \hspace{1.5cm}September 1992 \hspace{7.4cm}HLRZ-92-61\\
\begin{center}
\vspace*{0.5cm}
{{\large Chern-Simons term and   \\
Topological Charge on the Lattice}
 } \\
\vspace*{0.7cm}
{\large M.~L.~Laursen$^1$} \\
\vspace*{0.7cm}
{\normalsize
$\mbox{}^1$ {HLRZ, c/o KFA J\"{u}lich,
             P.O. Box 1913, D-5170 J\"{u}lich, Germany}}\\
\vspace*{2cm}
{\bf Abstract}
\end{center}
\setlength{\baselineskip}{1.3\baselineskip}

In a somewhat overlooked work by Seiberg, a definition of the
topological charge for SU(N) lattice fields was given. Here, it is
shown that Seibergs and L\"{u}schers charge definition are related
up to the section of the bundle.
With the continued interest in baryon number violating processes,
Seibergs paper is useful since it allows for a
Chern-Simons number also.
\newpage

\section{Introduction}
Over the past few years there has been quite an activity in the area
of baryon number violating processes.
Early on 't Hooft found that both baryon number and
lepton number are not conserved in the electroweak theory \cite{Hooft}.
While the $B-L$ symmetry remains unbroken
due to the anomaly cancellation, $B+L$ is no
longer conserved. This socalled baryon number violation is
caused by the nontrivial topological winding of the SU(2) gauge fields.
The anomaly of the fermionic current relates the winding of the gauge
fields and changes  the baryon number by an amount
\begin{equation}
 B(t_2) - B(t_1) = \frac{N_f}{16\pi^{2}}
                   \int_{t_1}^{t_2} \int d^{3}x
                    tr[F_{\mu\nu}\tilde{F}_{\mu\nu}]
\end{equation}
where $N_f$ is the number of families of quarks and leptons.
In the axial gauge $A_0=0$
one can relate the change in the baryon number
to the change in the Chern-Simons number
\begin{equation}
            B(t_2)-B(t_1) = N_{f}[N_{CS}(t_2)-N_{CS}(t_1)]
\end{equation}
where the Chern-Simons number  $N_{CS}$ is
\begin{equation}
           N_{CS}= - \frac{1}{8\pi^{2}} \int d^{3}x
   \epsilon_{ijk} tr[A_{i}(\partial_{j}A_{k}+\frac{2}{3}A_{j}A_{k})].
\end{equation}
At zero temperature such processes are exponentially suppressed as
$exp(-2\pi/\alpha_W)$, $\alpha \approx 1/30$. This is because any
gauge field configuration which changes the winding number has an action
at least that of the barrier height
$2\pi/\alpha_W$.
At high temperatures thermal fluctuations allow the system to tunnel
classically, since the only suppression factor is the
Boltzmann factor $exp(-\beta E/T)$ where $E$ is the barrier height  and
$T$ the temperature. As a consequence
any baryon asymmetry generated at the GUT scale will
get washed out as the universe approaches
the electroweak phase transition from above \cite{Kuzmin}.

Several lattice studies of baryon number violating
processes exist in the four dimensional SU(2) Higgs model
\cite{Ambjorn}. The configurations
are prepared at high temperature and the system is allowed
to change via the classical Hamiltonian equation of motion.
Since the axial gauge is used, the Gauss
constraint must be implemented in addition.
It is then possible to monitor
$\Delta N_{CS}$  during the time evolution
as a function of the temperature.
In practice  a
naive lattice definition of $F\tilde F$  is used to measure $Q$. Only
if the fields are very smooth is this method meaningful.
Notice, measuring $Q$
does not say anything about the value of $N_{CS}(t)$.
All these calculations are done in the real time formalism, and it
is worth while to study the behavior of $N_{CS}$ in
Euclidean time. Using a naive lattice transcription of this makes no
sense, since $N_{CS}$ must change by an integer under large gauge
transfomations.

There are now several (all geometric)
definitions for $N_{CS}$ in the Euclidean version  ref.~\cite{Luesch},
\cite{Gock}. For an actual calculation, the work of Seiberg
seems to be the most appropiate \cite{Karsch}.
In the work of G\"{o}ckeler et
al a continuum field   is constructed directly from the lattice field.
It is the purpose of this note to explain Seibergs lattice
Chern-Simons term and topological charge.
In particular I will show that Seibergs charge is related
to L\"{u}schers charge up to a section of the L\"{u}scher bundle.

\section{Topological charge and the Chern-Simons term in the continuum}
I  will first define the topological charge and I  start with the
SU(2) gauge field $A_{\mu}$ and     the gauge field tensor
$F_{\mu\nu}$:
\begin{equation}
 F_{\mu\nu} = \partial_{\mu}A_{\nu} - \partial_{\nu}A_{\mu}
      + [A_{\mu},A_{\nu}].
\end{equation}
Under a a local gauge transformation $g$  the gauge field changes as:
\begin{equation}
 \delta A_{\mu} =
 g^{-1}[A_{\mu} + \partial_{\mu}]g(x),
\end{equation}
while the gauge field tensor transforms    gauge covariantly
\begin{equation}
 F_{\mu\nu} \rightarrow
 g^{-1}F_{\mu\nu}g.
\end{equation}
The topological charge $Q$ is gauge invariant and an integer,
\begin{equation}
 Q = - \frac{1}{32\pi} \int_{M} d^{4}x
 \epsilon_{\mu\nu\rho\sigma}tr[F_{\mu\nu}F_{\rho\sigma}] \in Z.
\end{equation}
The manifold is denoted M and I  shall assume that its boundary
$\partial M$ is a three sphere $S^3$.
The topological charge density $q$ can be written
as  a perfect derivative
\begin{equation}
 q = - \frac{1}{32\pi}
 \epsilon_{\mu\nu\rho\sigma}tr[F_{\mu\nu}F_{\rho\sigma}]
        =  \partial_{\mu}K_{\mu}
\end{equation}
where the Chern-Simons density $K_{\mu}$ is
\begin{equation}
 K_{\mu} = - \frac{1}{8\pi^{2}} \epsilon_{\mu\nu\rho\sigma}
   tr[A_{\nu}(\partial_{\rho}A_{\sigma}+
   \frac{2}{3}A_{\rho}A_{\sigma})].
\end{equation}
It is gauge variant and changes under the gauge transformation $g$
by an amount
\begin{eqnarray}
 \delta K_{\mu} = & - & \frac{1}{24\pi^2}
  \epsilon_{\mu\nu\rho\sigma}tr[g^{-1}\partial_{\nu}g\,
 g^{-1}\partial_{\rho}g\,g^{-1}\partial_{\sigma}g] \nonumber \\
                                   & - & \frac{1}{8\pi^2}
  \epsilon_{\mu\nu\rho\sigma}
  \partial_{\nu}tr[\partial_{\rho}g\,g^{-1}A_{\sigma}].
\end{eqnarray}
I  define the Chern-Simons  number $N_{CS}$ in the axial gauge
as follows:
\begin{equation}
 N_{CS} = \int_{\partial M=S^3} d^{3}x K_{0} \not \in Z.
\end{equation}
While $N_{CS}$ is only an integer for pure gauge configurations,
the gauge variation is always an integer (the boundary term vanishes)
\begin{equation}
 \delta N_{CS} = - \frac{1}{24\pi^2}
          \int_{\partial M=S^3} d^{3}x
  \epsilon_{0\nu\rho\sigma}tr[g^{-1}\partial_{\nu}g\,
 g^{-1}\partial_{\rho}g\,g^{-1}\partial_{\sigma}g] \in Z.
\end{equation}
This becomes clear from homotopy theory using the mapping
$g: S^{3} \rightarrow SU(2) = S^{3}$.
Such mappings are characterized with the homotopy class
$\Pi_{3}(S^{3}) \in Z$.

\section{Topological charge and the Chern-Simons term on the lattice}
I will now consider the lattice version  of the topological charge $Q$
and the  Chern-Simons number $N_{CS}$.
I shall first follow L\"{u}schers geometric method to define $Q$.
Problems with dislocations will be ignored here.
Let the manifold be a four torus
 $M = T^4$ and assume it is covered with cells (hypercubes) $c(n)$.
Consider the  gauge potential
$A_{\nu}^{n}$ defined on $c(n)$ and likewise let
$A_{\nu}^{n-\hat{\mu}}$ be defined on $c(n-\hat{\mu})$.
At the faces  (cubes)
$f(n,\mu) = c(n-\hat{\mu}) \cap c(n)$, one    relates
the two potentials by   a  transition function
$v_{n,\mu}$
\begin{equation}
  A_{\nu}^{n-\hat{\mu}}  =
  v^{-1}_{n,\mu} [A_{\nu}^{n} + \partial_{\nu}] v_{n,\mu}.
\end{equation}
L\"{u}scher now fixes to a local complete axial gauge
in each $c(n)$. For any corner $x$ of $c(n)$ one defines a parallel
transporter $w_{n}(x)$ from  $n$ to $x$. This leads to
$v_{n,\mu}(x) = w_{n-\hat \mu}(x)\, w_{n}^{-1}(x)$.
It is then possible to interpolate it to the whole
cube, thus defining a bundle. One finds
$v_{n,\mu}(x) = s^{n-\hat \mu}_{n,\mu}(x)^{-1}\,
                     v_{n,\mu}(n)\, s^{n}_{n,\mu}(x)$.
Here, the function $s$ is defined on $f(n,\mu )$.
Let $p$ be the restriction of $s$ to $\partial f(n,\mu )$.
The actual expressions are given in ref.~\cite{Luesch}.
The topological charge is:
\begin{equation}
Q^{L}   =  \sum_{n} q^{L}(n)  =  \frac{1}{2\pi}
             \sum_{n,\mu}(-1)^{\mu}(k_{n,\mu} - k_{n+\mu,\mu}),
\end{equation}
where
\begin{eqnarray}
 (-1)^{\mu} k_{n,\mu} = &  &
 \frac{1}{12\pi} \epsilon_{\mu\nu\rho\sigma} \int_{f} d^{3}x
  tr[s\partial_{\nu}s^{-1}\,
     s\partial_{\rho}s^{-1}\,
     s\partial_{\sigma}s^{-1}] \nonumber \\
     +  &     &
 \frac{1}{4\pi} \epsilon_{\mu\nu\rho\sigma} \int_{\partial f} d^{2}x
  tr[p^{-1}\partial_{\rho}p\,
     s^{-1}\partial_{\sigma}s].
\end{eqnarray}
At this point one should note that $q^{L}(n)$ is gauge invariant
and unrestricted.
In Seibergs version no local gauge fixing is performed, but otherwise
the same interpolation is performed. Replace
$(s,p,k_{n,\mu}) \rightarrow  (S,P,K_{n,\mu})$. The difference is that
$S$ and $P$ only depend on the original gauge fields in the cube.
Then
\begin{equation}
 N_{CS} = \frac{1}{2\pi} \sum_{n_s} K_{n_{s},\mu}
\end{equation}
is nothing but a Chern-Simons term
(the summation is over the spatial lattice only).
Like in the continuum
it is only an integer for pure gauge field configurations, but under
gauge transformations it changes by an integer. The
corresponding topological charge is defined as
\begin{equation}
 Q^{S}   =  \sum_{n} \tilde q^{S}(n),\;\;\;
       -1/2 \leq \tilde q^{S}(n) < 1/2.
\end{equation}
By restricting the charge to this interval one has a gauge invariant
charge definition.
After some algebra the Chern-Simons term is
found to have the correct naive continuum limit, that is after writing
$U_{n,\mu} = exp(a\, A_{n,\mu} )$ and letting  $a \rightarrow 0$:
\begin{equation}
 (-1)^{\mu} K_{n,\mu} =
 \frac{a^3}{4\pi} \epsilon_{\mu\nu\rho\sigma}
   [A_{n,\nu}(\partial_{\rho}A_{n,\sigma}+
   \frac{2}{3}A_{n,\rho}A_{n,\sigma})].
\end{equation}

As an interesting corollary I find that the two charge definitions
are related. I first introduce the section of the L\"{u}scher bundle,
$w(x)$, $x \in \partial c(n)$, which relates $s$ and $S$:
$s(x) = w(0)S(x)w^{-1}(x)$ \cite{Goc1}. The following identity is useful
\begin{eqnarray}
 &  &\epsilon_{\mu\nu\rho\sigma}
  tr[(sw)\partial_{\nu}(sw)^{-1}\,
     (sw)\partial_{\rho}(sw)^{-1}\,
     (sw)\partial_{\sigma}(sw)^{-1}] \nonumber \\
 & = &\epsilon_{\mu\nu\rho\sigma}
  tr[s\partial_{\nu}s^{-1}\,
     s\partial_{\rho}s^{-1}\,
     s\partial_{\sigma}s^{-1}] \nonumber \\
 & - &\epsilon_{\mu\nu\rho\sigma}
  tr[w^{-1}\partial_{\nu}w\,
     w^{-1}\partial_{\rho}w\,
     w^{-1}\partial_{\sigma}w] \nonumber \\
 & + &
 3 \epsilon_{\mu\nu\rho\sigma}
  tr[w\partial_{\rho}w^{-1}\, s^{-1}\partial_{\sigma}s].
\end{eqnarray}
This gives
\begin{eqnarray}
 (-1)^{\mu} K_{n,\mu} = (-1)^{\mu} k_{n,\mu}
   & - &
 \frac{1}{12\pi} \epsilon_{\mu\nu\rho\sigma} \int_{f} d^{3}x
  tr[w^{-1}\partial_{\nu}w\,
     w^{-1}\partial_{\rho}w\,
     w^{-1}\partial_{\sigma}w] \nonumber \\
        & +   &
 \frac{1}{4\pi} \epsilon_{\mu\nu\rho\sigma} \int_{\partial f} d^{2}x
  tr[P^{-1}\partial_{\rho}P\,
     w^{-1}\partial_{\sigma}w].
\end{eqnarray}
Using the identity
\begin{eqnarray}
 \sum_{\mu ,\nu ,\rho ,\sigma}
 \epsilon_{\mu\nu\rho\sigma} \int_{p(n,\mu\nu )} d^{2}x
  tr[P^{-1}\partial_{\rho}P\,
     w^{-1}\partial_{\sigma}w] = 0,
\end{eqnarray}
and others I arrive at
\begin{eqnarray}
q^{S}(n) & = & q^{L}(n) - q^{w}(n) \nonumber \\
         & = & q^{L}(n) -
 \frac{1}{24\pi ^2} \epsilon_{\mu\nu\rho\sigma}
 \int_{\partial c(n)} d^{3}x
  tr[w^{-1}\partial_{\nu}w\,
     w^{-1}\partial_{\rho}w\,
     w^{-1}\partial_{\sigma}w],
\end{eqnarray}
where
$q^{w}(n)$
is the topological charge (integer) of the section.
Since $q^{w}(n)$  is not gauge invariant, the same is true for
$q^{S}(n)$.
It can change by an integer under a gauge
transformation. This is why one must use $\tilde q^{S}(n)$ instead of
$q^{S}(n)$. Notice also that
$Q^{S}   =  \sum_{n} q^{S}(n) = 0$ since
$Q^{L}   =  \sum_{n} q^{L}(n) =  \sum_{n} q^{w}(n)$.
Therefore $\tilde q^{S}(n) = q^{L}(n)$ up to integers.
For smooth fields like instantons they always agree, while for realistic
configurations this is true for almost every hypercube. It can happen
in a few hypercubes that $|q^{L}(n)| > 1/2$ and this often make the
charges different.
Since Seiberg interpolates the original gauge fields, he has a much
rougher function to integrate. Therefore it is necessary to perform
a global Landau  or axial gauge fixing, to make the integrals converge.
For the topological charge this is fine since it is gauge invariant. The
same is true for the restricted Chern-Simons term $\tilde N_{CS}$.

As a check I have taken an instanton configuration.
For the evaluation of the integrals see ref.~\cite{Karsch}.
The size of the lattice is $8^{3}\times 12$ and the
core size is $\rho = 2$. For both charge definitions I find $Q = -1$.
In Fig.~1 I have plotted the Chern-Simons number as a function
of the time slice. The   complete axial gauge was chosen to resemble
a tunneling event. Notice that $N_{CS} = 0$ in the first
time slice and  $N_{CS} = 1$ in the next to last time slice. Of course,
at the last time slice the Chern-Simons number must return to it's
initial value due to periodicity.
\newline
\noindent
{\large \bf Acknowledgement}

\noindent
I appreciate valuable discussions with F.~Karsch and B.~Plache.
\newpage

\newpage
\noindent
{\large \bf Figure captions}

\noindent
Figure 1.
Profile of $N_{CS}$ through an instanton configuration. The lattice
is $8^{3}\times 12$ and the core size $\rho = 2$.

\end{document}